\documentclass[runningheads]{llncs}
\usepackage{graphicx}
\usepackage{amsmath}
\usepackage{caption}
\usepackage{subcaption}

\usepackage[misc,geometry]{ifsym} 
\usepackage{todonotes}
\usepackage[ruled,vlined]{algorithm2e}
\usepackage{hyperref}

\begin{document}
\title{An Interpretable Approach to Automated Severity Scoring in Pelvic Trauma}
\author{Anna Zapaishchykova\inst{1,2}
\and David Dreizin\inst{3} \and
Zhaoshuo Li\inst{1} \and
Jie Ying Wu\inst{1}\and
Shahrooz Faghih Roohi\inst{2} \and
Mathias Unberath({\Letter})\inst{1}
}
\authorrunning{Zapaishchykova et al.}

\institute{Johns Hopkins University, USA\\
\email{mathias@jhu.edu}
\and Technical University of Munich, Germany
\and University of Maryland, School of Medicine, USA}
\maketitle

\begin{abstract}
Pelvic ring disruptions result from blunt injury mechanisms and are often found in patients with multi-system trauma. To grade pelvic fracture severity in trauma victims based on whole-body CT, the Tile AO/OTA classification is frequently used. 
Due to the high volume of whole-body trauma CTs generated in busy trauma centers, an automated approach to Tile classification would provide substantial value, e.\,g., to prioritize the reading queue of the attending trauma radiologist. In such scenario, an automated method should perform grading based on a transparent process and based on interpretable features to enable interaction with human readers and lower their workload by offering insights from a first automated read of the scan. This paper introduces an automated yet interpretable pelvic trauma decision support system to assist radiologists in fracture detection and Tile grade classification. The method operates similarly to human interpretation of CT scans and first detects distinct pelvic fractures on CT with high specificity using a Faster-RCNN model that are then interpreted using a structural causal model based on clinical best practices to infer an initial Tile grade. The Bayesian causal model and finally, the object detector are then queried for likely co-occurring fractures that may have been rejected initially due to the highly specific operating point of the detector, resulting in an updated list of detected fractures and corresponding final Tile grade. Our method is transparent in that it provides finding location and type using the object detector, as well as information on important counterfactuals that would invalidate the system's recommendation and achieves an AUC of 83.3\%/85.1\% for translational/rotational instability. Despite being designed for human-machine teaming, our approach does not compromise on performance compared to previous black-box approaches.

\end{abstract}

\section{Introduction}
Mortality in patients with pelvic fractures ranges from 4\% to 15\%~\cite{Vaidya2015Dec} depending on fracture severity. Urgent surgical or endovascular treatment is necessary to ensure good patient outcome, however, pelvic ring disruption severity grading remains a subjective bottleneck in patient triage. In contemporary practice, whole-body CT is routinely performed upon admission of trauma victims with suspected pelvic fractures~\cite{Coccolini2017Jan,Dreizin2012May}, which are then commonly graded with the unified AO Foundation/Orthopaedic Trauma Association (AO/OTA) system~\cite{TileM.1996}. CT scans are interpreted by readers with widely disparate levels of experience in a time-sensitive environment resulting in limited accuracy and inter-observer agreement. An objective, automated approach to Tile AO/OTA grading of pelvic fractures could harmonize the accuracy of Tile grading across readers while accelerating the interpretation of whole-body CT to extract pelvic fracture-related risk scores, thus addressing an important unmet clinical need~\cite{Vaidya2015Dec}. Previous work using automated deep learning methods to evaluate pelvic fracture severity relied on black-box models that predict severity from trauma imagery~\cite{dreizin2021automated,sato2020computer} but do not provide a transparent, interpretable association between Tile AO/OTA grade and detectable fractures. Plain radiographs have been used for pelvic fracture detection~\cite{Cheng2021Feb}, but not classification tasks, and are severely hampered by bone and soft tissue overlap. CT remains the imaging gold standard.

Irrespective of the imaging modality used for fracture grading, black-box approaches are undesirable in a high-stakes decision making because they are designed to be run as a standalone application. They do not offer immediate opportunity for interaction and may be prone to automation bias. 
Interpretable models that offer insights into the decision-making process may promote human-machine teaming because they enable human interaction and verification, thus potentially bolstering clinicians' confidence in the recommendation. Prior methods~\cite{chang2018explaining,fong2019understanding,lenis2020domain} have been proposed to fill this gap by explaining causality through the identification of regions of interest that contribute to a predictor’s outcome.

\subsubsection{Contribution} We present an automated yet interpretable algorithm for first-order Tile AO/OTA grading from trauma CT. The system mimics a human expert read of the CT scan by first identifying fracture types and locations via a deep object detection algorithm, and then inferring Tile grade from presence or absence of fractures using a Bayesian network. An overview of the method is provided in Fig.~\ref{method}. By establishing causal relationships between Tile grade and fracture presence, the framework provides a transparent inference pipeline that supplies fracture location and type, as well as information on counterfactuals, e.\,g. missed or misclassified fractures, that would invalidate the system's recommendation. As such, the system is designed to decrease the workload of the attending radiologist by facilitating validation and refinement.

\section{Related Work}
Specific fracture patterns and abnormal bony relationships in different parts of the pelvic ring on CT correspond with the degree of mechanical instability~\cite{TileM.1996}. This section outlines pelvic fracture biomechanics and briefly reviews prior work on fracture detection and interpretability for medical decision making.

\subsubsection{Fracture Detection on Plain Radiographs and CT}
Previous studies have demonstrated the feasibility of deep learning for fracture detection in various anatomical regions from plain radiographs~\cite{Cheng2021Feb,kalmet2020deep}. Developments in CT fracture detection have been presented for the rib cage~\cite{Blum2021Jan}, spine~\cite{Burns2020Jan} and skull~\cite{kalmet2020deep}.  
We use Faster-RCNN~\cite{Ren2015Jun} for fracture localization and classification to extract pelvic ring disruptions from CT scans since it is a well-established architecture for both general purpose object and fracture detection~\cite{abbas2020lower,Yahalomi2018Dec}.

\subsubsection{Features of pelvic ring disruption}
The AO/OTA have adopted the Tile classification scheme for grading mechanical instability and severity of pelvic fractures~\cite{Tile1988Jan}. First-order Tile grading (Grade \textbf{A} - translationally(T) and rotationally(R) stable, Grade \textbf{B} - T stable and R unstable, Grade \textbf{C} - both T and R \textit{i.e. "globally"} unstable) has been shown to correlate with major arterial injury, need for surgical packing or catheter-based interventions, need for massive transfusion, and mortality~\cite{dreizin2019commentary,Dreizin2018Jun}. Different degrees of instability manifest with different types of abnormal bony relationships on the CT scan, including pubic symphysis diastasis (PSD), divergent or parallel sacroischial (SI) joint diastasis, non-diastatic and diastatic sacral fractures, ischial spine avulsions (ISp), and innominate bone fractures involving the anterior and posterior pelvic ring~\cite{Dreizin2016Sep,TileM.1996}. Fragility in elderly patients is not expressly included in Tile's framework, but has come to be recognized as an important modifier of fracture severity~\cite{Vaidya2016Mar}.

\subsubsection{Interpretable Artificial Intelligence for Medical Decision Support}

There is increasing evidence that medical decision support systems that are interpretable or explainable are perceived as more trustworthy~\cite{ahmad2018interpretable,Scholkopf2021Feb}. One way of making such models interpretable is to derive predictions with a task-specific causal model that is created from domain expertise and can model uncertainties between variables and the final prediction~\cite{tonekaboni2019clinicians}. Using directed graphical models to encode the causal structure between variables~\cite{pearl2014probabilistic}, a complete causal graph can be constructed to induce the reasoning around diagnostic predictions~\cite{Castro2020Jul}. 
Here, we combine the benefits of Bayesian structural causal models and deep learning-based feature extractors to create an automatic yet interpretable algorithm for pelvic Tile grading from CT scans. 

\begin{figure}[htpb]
\centering
\includegraphics[width=0.75\textwidth]{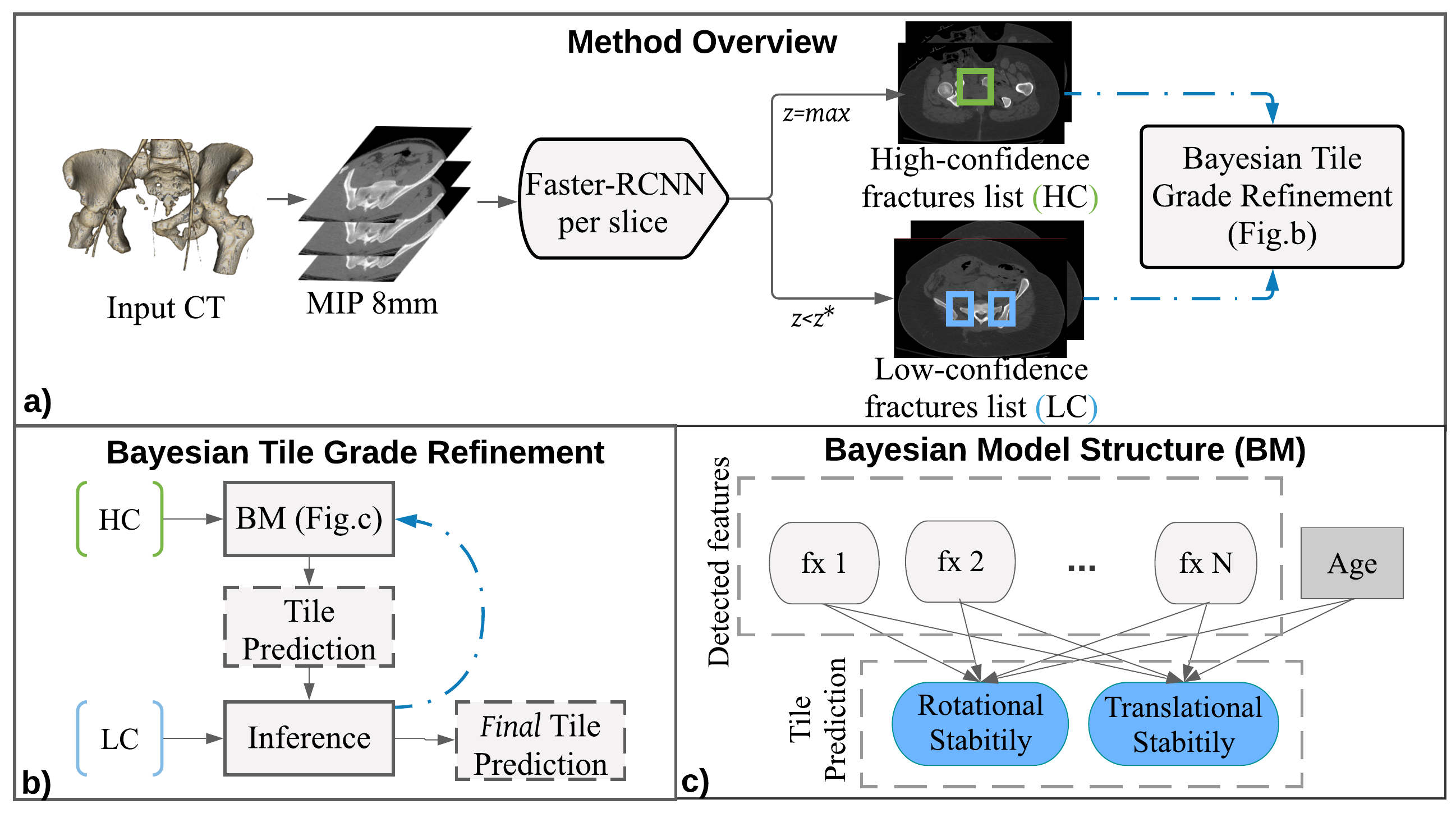}
\caption{a) Method overview. After training on maximum intensity projections (MIPs) and corresponding annotations generated from 3D CT segmentation maps of key fracture classes that contribute to Tile grading, Faster-RCNN is used to extract high- and lower-confidence findings, such as fractures. b) Tile grade refinement algorithm using causal Bayesian model. Detailed description can be found in Alg.~\ref{alg:BM}. c) Structure of the Bayesian model (BM): [fx 1 - fx N] are findings detected by the Faster-RCNN. Tile grade is represented by a combination of translational and rotational instability.} 
\label{method}
\end{figure}

\section{Methods}
To detect fractures and other pertinent findings (for brevity, we will summarize all of such findings under the term \emph{fractures}) in a pelvic CT volume, we use a Faster-RCNN trained on annotated bounding boxes on axial maximum intensity projections (MIP) slices (Fig.~\ref{method}a). After fracture locations and fracture types have been determined together with their corresponding confidence score, we obtain an initial Tile grade estimate from the Bayesian model (BM) applied to the detected fractures with high confidence (see Fig.~\ref{method}b). We include the patient's age, which helps in clinical practice to determine overall severity because fractures tend to propagate through the weaker bones in the elderly (see Fig.~\ref{method}c for structure of the BM). By combining patient-specific information with fractures found by Faster-RCNN, we can compute inferences conditioned on the fractures that were detected with high confidence. Due to the highly specific operating point selected for Faster-RCNN, it is possible that some true fractures were erroneously rejected. To refine the Tile grade estimate based on these erroneously rejected fractures and simultaneously improve fracture detection, we intervene on the initial detected fractures to identify potential misses that would confirm or invalidate the current Tile estimate. We then check for presence of these candidate fractures identified by our causal model among the lower confidence fractures, and if present, include them in our high-confidence findings. Finally, we run our BM on the updated list of fractures to obtain the binding estimate of Tile grade. 

\begin{algorithm}[htpb]
\caption{Proposed Tile refinement algorithm using a Bayesian network structural causal model (BM)}
\label{alg:BM}
\SetAlgoLined
BM$(x)$ is a method for computing the marginal likelihood for nodes given $x$\;
$p(x)$ is a probability of a variable $x$\;
\textbf{Input:}\\
$z$ is a confidence threshold setting\;
$FX_{high}$ is list of fractures detected with high confidence $>z*$\;
$FX_{low}$ is list of fractures detected with low confidence $<z*$\;
\textbf{Result:} Tile grade prediction - A,B or C\;
  Compute BM on $FX_{high}$ to get initial Tile grade\;
  \For{$Tile$ \textbf{in} [A,B,C]}
  {
  Compute BM$(FX_{high},Tile)$ to get the $p$ for features(\textbf{$fx$})\;
    \If{\textbf{$fx$} has $p>z*$ and is in $FX_{low}$}
    {
    Update $FX_{high}$ by adding $fx$\;
    }
  }
  Compute BM on updated $FX_{high}$\;
  argmax ($p(Tile)$)\;
\end{algorithm}

The subset of pelvic fracture (fx)-related features that were determined a priori to be clinically meaningful for Tile grading by trauma radiologists includes: PSD; anteriorly divergent SI and parallel SI joint diastasis; non-diastatic and
diastatic sacral fx; ISp, innominate bone fractures involving the anterior and posterior pelvic ring (henceforth "ring fx").

The proposed algorithm is formally described in Alg.~\ref{alg:BM}. An initial estimate of the Tile grade will be made by selecting the high-confidence features $FX_{high}$ of pelvic ring disruption with confidence threshold set to $z=max$ to range [0.95-1.0], so that the feature detection network will have high specificity. By computing the marginal likelihood probabilities for each possible Tile state, we get a list of potentially missed fracture labels that are more likely to co-occur with the highly specific conditions found in the specific CT. For every fracture in this candidate list, we check its presence in a list of lower-confidence fracture detections $FX_{low}$ that is generated using a lower threshold $z<z*$. If such matching between the $FX_{low}$ and fractures suggested by BM are found, $FX_{high}$ will be updated by considering the additional features found in $FX_{low}$, and the predicted Tile grade will be updated based on this new feature set.

While we consider our algorithm from a fully automated perspective in this manuscript, the refinement procedure based on a structural causal model is designed with immediate benefit for clinical workflows in mind. This is because, instead of setting a different threshold $z$ for possibly missed fractures, our model could alert the attending radiologist for review of the fractures. Similarly, the BM could calculate counterfactuals to the current prediction to identify possible misses that would change treatment recommendations. Because the object detection network supplies bounding boxes for each fracture finding, our method allows for effective visualization of all findings.This may contribute to an effective hand-over from an autonomous to human-machine teaming operation.

\section{Experiments and Results}

\subsection{Dataset and Metrics}
We obtained approval for use of the dataset initially described in~\cite{dreizin2021automated}, that consists of 373 admission CT scans of adult patients from two level I trauma centers with bleeding pelvic fractures and varying degrees of instability with IRB-approved waiver of consent.
Imaging was performed with intravenous contrast in the arterial phase on 40-, 64-, and dual source 128-row CT scanners and reconstructed with 0.7 – 5\,mm slice thickness. Patient-level annotation using the Tile AO/OTA grading system was performed by three trauma-subspecialized radiologists and the consensus vote was used as gold standard for training and evaluation. Presence and location of fractures and fracture-related features such as pelvic ring disruption were performed on the 3D CTs by one radiologist with trauma expertise at the primary study site.

\subsection{Implementation Details}

The CT scans were interpolated, isotropically scaled and normalized on a per-volume basis. To preserve some volumetric information while reducing the complexity of the problem from 3D to 2D, MIP was used to generate 8\,mm-thick slices along the axial plane. Reference standard segmentation of pelvic fractures were propagated to these MIPs as bounding boxes with corresponding fracture label. On all MIPs containing labels, we train a Faster-RCNN using TensorFlow for 4000 epochs with a ResNet-50 backbone pretrained on ImageNet to extract pelvic features.
For evaluation, we used five-fold cross validation where, in each fold, 80\% of the data was used for training and 20\% of the data was used for testing. Random flip and random rotations were used for data augmentation. To reduce the number of false positives, we gather poorly predicted features during training and up-weigh them in the next training phase for 3000 epochs. The structure of BM is illustrated in Fig.~\ref{method}c. The BM is developed using the CausalNex~\cite{BibEntry2021Feb} package. We fit probabilities using Bayesian parameter estimation and learn conditional probability distributions (CPDs) w.r.t. Tile expert consensus from the same data split used during the fracture detection training.

\subsection{Experimental Results and Discussion}

To evaluate the Tile AO/OTA assessment, we split the Tile grade into component instabilities (\textit{rotational} or \textit{translational}), as in~\cite{Dreizin2016Sep,Dreizin2018Jun}. Human readers explicitly search for evidence of rotational and translational instability separately to synthesize the Tile grade and this represents the most intuitive real-world approach to the problem. We present all scores as an average over the 5 folds (Tab.~\ref{tab:auc}) and use area under the receiver operating characteristic curve (AUC) for classification performance and Kappa scores~\cite{McHugh2012Oct} for reader agreement measurements. Inter-observer reliability between automated predictions and expert consensus improved using the proposed refinement technique to a level that is close to the inter-rater variability of human observers~\cite{zingg2020interobserver}. Since inter-observer reliability is only moderate to fair for many Tile grades~\cite{zingg2020interobserver}, we hypothesize the lack of agreement is responsible for the imperfect performance of \textbf{BM($GT$)} on expert-annotated fractures. Tab.~\ref{tab:auc} shows the full results.

\begin{table}[h]
  \centering
  \caption{Average AUC and Kappa scores for rotational (R) and translational (T) instabilities. All columns are compared to the patient-level expert consensus Tile grade by three radiologists.  \textbf{BM($GT$)} represents Bayesian model (BM) fitness when using the reference standard fracture detections used to train the Faster-RCNN model; \textbf{BM($FX_{low}$)} represents BM prediction on automatically detected fractures when lower confidence fractures (with $z=0.5$) are immediately included in Tile grade inference; \textbf{BM($FX_{high}$)} represents BM prediction on automatically detected fractures with high confidence ($z=max$); \textbf{BM refinement} represents predictions after the proposed refinement pipeline. }\label{tab:auc}
  \resizebox{\textwidth}{!}{
      \begin{tabular}{|p{1.3cm} | p{1.3cm} |  p{1.3cm} || p{1cm} | p{1cm} | p{1cm} | p{1cm}|| p{1.3cm}  | p{1.3cm} |}
        \hline
        \multicolumn{1}{|p{1.3cm}|}{} &
        \multicolumn{2}{p{2.6cm}||}{BM($GT$)} &
        \multicolumn{2}{p{2cm}|}{BM($FX_{low}$)}  &
        \multicolumn{2}{|p{2cm}||}{BM($FX_{high}$)} & \multicolumn{2}{p{2.6cm}|}{\textbf{BM refinement}}  \\
        \hline
         Metrics & R & T & R & T & R & T & R & T \\
        \hline
        AUC & 0.81 & 0.84 & 0.71 & 0.63 & 0.78 & 0.82 & \textbf{0.85} & \textbf{0.83}  \\
        \hline
        Kappa & 0.32 & 0.48 & 0.1 & 0.18 & 0.15 & 0.38 & \textbf{0.24} & \textbf{0.5}  \\
        \hline
      \end{tabular}
  }
\end{table}

\begin{figure}[h]
\centering
\includegraphics[width=0.8\textwidth]{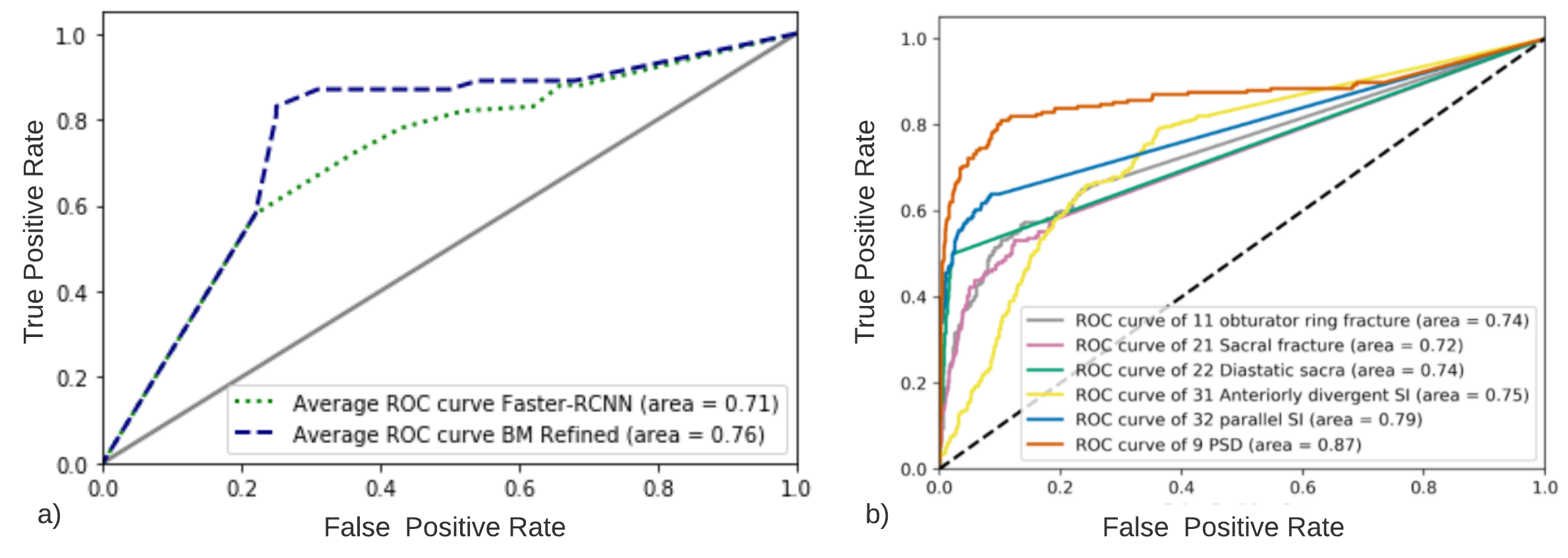}
\caption{a) Average RoC curve of Faster-RCNN-based fracture detection with and without BM-based refinement for $z=0.5$ b) RoC curves for every fracture type using Faster-RCNN at $z=0.5$} \label{fig:roc}
\end{figure}

We evaluated Faster-RCNN for each fracture type separately (Fig.~\ref{fig:roc}b). We then evaluate the performance of Faster-RCNN for fracture detection with and without BM-based refinement(Fig.~\ref{fig:roc}a) and show that the proposed scheme not only improves Tile grading but also positively affects fracture detection when the initial network operates at high specificity. 

We found that the fracture with the lowest confidence by the object detector (Anteriorly Divergent SI) is often responsible for the Tile prediction update. Even though they were initially left out, anteriorly divergent SI fractures were included after the refinement. In a situation where a binder (a device used to stabilize suspected pelvic fractures in patients with traumatic injury) is present, the subset of visible fractures will appear less severe, even though presence of Anteriorly Divergent SI indicates a major high energy blunt trauma event. An example of a case is shown in the supplementary material.

\begin{figure}
\centering
\includegraphics[width=0.8\textwidth]{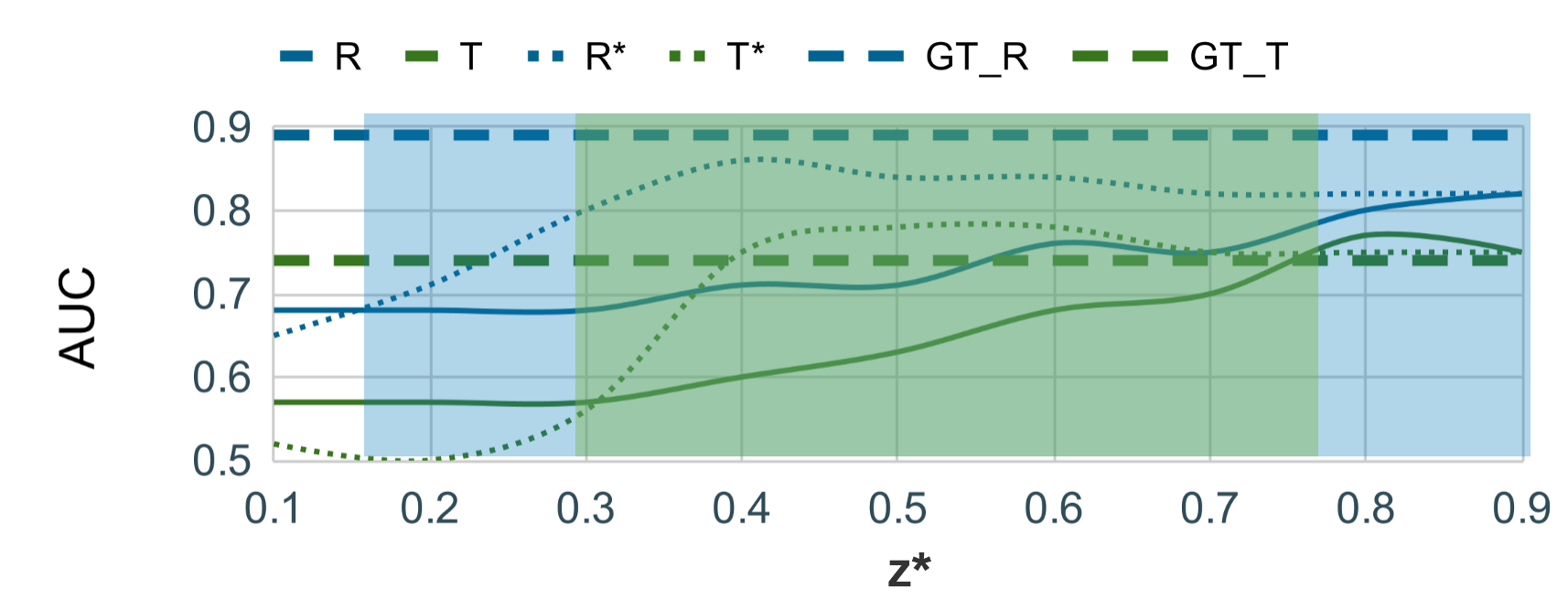}
\caption{Impact of the magnitude of $z$ of $FX_{low}$ on the AUC of Tile prediction for $FX_{high}(z=0.95)$. \textbf{R/T} is BM model prediction without refinement on a list of fractures with confidence $>z*$, \textbf{R*/T*} is BM model prediction after refinement, \textbf{GT\_R/GT\_T} is BM model prediction on the ground truth semantic labels. The proposed method outperforms the BM without refinement when $z*\in[0.3-0.75]$ and therefore is less dependent on the initial parameter selection.} \label{fig:func_from_lc}
\end{figure}

To measure the impact of the confidence threshold $z$ of the lower-confidence detections on Tile prediction, we compare the performance of BM on predicting Rotational (R)/Translational (T) instabilities based on fracture detections $FX_{low}$ achieved with a Faster-RCNN operating at a confidence threshold in an interval $z\in[0.1-0.9]$, while $z=0.95$ is held constant for the high-confidence detections (Fig.~\ref{fig:func_from_lc}). We find that the proposed refinement method improves upon direct inference on a less specific detection result without being overly dependent on hyperparameter selection. 

\section{Conclusion}

In this work, by leveraging the respective strengths of deep learning for feature detection and a BM for causal inference, we achieve automatic and interpretable Tile grading of pelvic fractures from CT images. Using the proposed pipeline, Tile grade prediction had an AUC of 83.3\% and 85.1\% for translational and rotational instability comparable to the previous black-box method~\cite{dreizin2021automated} but with added interpretability and potential for human-machine teaming.

The described methodology can benefit from interaction with experienced
radiologists in place of the autonomous refinement, to further increase the inter-observer agreement. One limitation to be addressed in the future, either algorithmically or through interaction, are false-false detections, i.e., false positives that are detected with high confidence and then propagated unchanged through the refinement step. Because our BM-based refinement strategy cannot currently reject high-confidence fractures, these findings would irrevocably bias the Tile grade estimation. 
Another line of future work will include investigations on the effectiveness of the proposed system when working with attending trauma radiologists in a collaborative setting to see if the system can indeed decrease the time to triage. We anticipate that the core ideas of the presented framework may find application across various diagnostic image interpretation tasks. We understand our methods as a step towards machine-human teaming for quick and accurate pelvic fracture severity scoring using the Tile AO/OTA system to standardize and accelerate triage, and ultimately, reduce mortality rates. 

\newpage

\bibliographystyle{splncs04}
\bibliography{reference}

\newpage

\end{document}